# Radon induced hyperplasia: effective adaptation reducing the local doses in the bronchial epithelium

Balázs G. Madas

*Centre for Energy Research, Hungarian Academy of Sciences*
*Environmental Physics Department*
*1121 Budapest, Konkoly-Thege Miklós út 29-33., Hungary*

*Tel.: +36 1 3922222, ext.: 1981*
*Fax: +36 1 3922712*
*E-mail: balazs.madas@energia.mta.hu*

**Abstract.** There is experimental and histological evidence that chronic irritation and cell death may cause hyperplasia in the exposed tissue. As the heterogeneous deposition of inhaled radon progeny results in high local doses at the peak of the bronchial bifurcations, it was proposed earlier that hyperplasia occurs in these deposition hot spots upon chronic radon exposure. The objective of the present study is to quantify how the induction of basal cell hyperplasia modulates the microdosimetric consequences of a given radon exposure. For this purpose, numerical epithelium models were generated with spherical cell nuclei of six different cell types based on histological data. Basal cell hyperplasia was modelled by epithelium models with additional basal cells and increased epithelium thickness. Microdosimetry for alpha-particles was performed by an own-developed Monte-Carlo code. Results show that the average tissue dose, and the average hit number and dose of basal cells decrease by the increase of the measure of hyperplasia. Hit and dose distribution reveal that the induction of hyperplasia may result in a basal cell pool which is shielded from alpha radiation. It highlights that the exposure history affects the microdosimetric consequences of a present exposure, while the biological and health effects may also depend on previous exposures. The induction of hyperplasia can be considered as a radioadaptive response at the tissue level. Such an adaptation of the tissue challenges the validity of the application of the dose dose rate effectiveness factor from a mechanistic point of view. As the location of radiosensitive target cells may change due to previous exposures, dosimetry models considering the tissue geometry characteristic of normal conditions may be inappropriate for dose estimation in case of protracted exposures. As internal exposures are frequently chronic, such changes in tissue geometry may be highly relevant for other incorporated radionuclides.

## Introduction

The health effects of radon exposure are of significant importance both to public health and in radiological protection. Radon is considered as the second most important cause of lung cancer after smoking (NRC, 1999). Its contribution to the effective dose originating from the natural background radiation is approximately 50 %, and its contribution to the effective dose from all background radiation exposure is about 42 % (UNSCEAR, 2000). However, the mechanisms leading from radon exposure to lung cancer are quite unclear.

The heterogeneous aerosol deposition results in heterogeneous spatial activity distribution of inhaled radon progeny in the bronchial airways (Baláshazy et al. 2009). Due to the short range of emitted alpha-particles and the relatively short half life of radon progeny, the heterogeneity is reflected in the dose distributions too (Madas et al. 2011). Whereas it remains an open question whether the measure of dose heterogeneity increases or decreases the risk, the understanding of the mechanisms leading from radon exposure to lung cancer is not expected if the effects of locally high doses are not studied (Madas 2016).

Alpha-particles effectively kill cells (Hei et al. 1997; Soyland and Hassfjell 2000) resulting in high cell death rate in the deposition hot spots (Madas and Baláshazy 2011). In order to maintain tissue homeostasis and function, the replacement of lost cells is necessary, i.e. the total number of cell divisions has to be increased in case of increased cell death rate. In general, there are two ways for enhancing the total number of cell divisions: the increase of the division rate of progenitor cells and the increase in the number of progenitor cells, i.e. hyperplasia (Lander et al. 2009; Marciniak-Czochra et al. 2009). Theoretical considerations show that increasing the number of progenitor cells is a more effective way for tissue regeneration than increasing the cell division rate, although the simultaneous increase of both is the most effective (Lander et al. 2009; Marciniak-Czochra et al. 2009). With regard to the effects of ionizing radiation, changes in tissue architecture were modelled in case of the haematopoietic system and small intestine (Smirnova 2009).

Besides the theoretical considerations, the potential induction of hyperplasia in the lungs is also supported by experimental (McDowell et al. 1979) and histological data (Auerbach et al. 1961). The cell type of origin may be different even within the lungs. Chronic irritation and stimulation results in hyperplasia of progenitor basal cells residing in the respiratory epithelium (Gordon et al. 2009), while goblet cells also increase in number in response to a wide variety of drugs and irritants (Rogers 1994; 2003), and cigarette smoke and nicotine cause neuroendocrine cell hyperplasia in hamsters (Tabassian et al. 1989). It is important to note that exposures just over periods of a few days to weeks to different irritants markedly increases goblet cell number in the more proximal airways and induces the appearance of goblet cells in the more distal airways (Rogers 1994).

On the basis of these observations and theoretical considerations, it was put forward earlier that progenitor cell hyperplasia occurs in the deposition hot spots upon chronic cell death caused by radon progeny (Madas and Baláshazy 2011). However, the increase in the number of cells implies an increase in the epithelial thickness (Rogers 1994) along with other changes in the tissue geometry. As the tissue geometry and in particular the location of radiosensitive target cells are major determinant of effective dose of inhaled radionuclides (ICRP, 1994), the consequences of such changes in tissue geometry are of interest to radiological protection.

The objective of the present study is to quantify the microdosimetric consequences of hyperplasia induced by radon exposure. Cellular burdens like hits received and dose absorbed by progenitor cells will be determined as the function of alpha-decays per unit surface. While the induction of hyperplasia was proposed only above a threshold dose (Madas and Baláshazy 2011), theoretical considerations (Lander et al. 2009; Marciniak-Czochra et al. 2009) suggest that hyperplasia may occur even at lower dose rates. Nevertheless, it is expected that higher dose rates from alpha-particles results in higher measure of hyperplasia, and therefore, the focus of the study is on the deposition hot spots.

## Methods

*Generation of numerical epithelium models*

In order to determine hits received and doses absorbed by cell nuclei, numerical models of the epithelium of the large bronchi were developed. As only cell nuclei are considered, the model is much simpler than our earlier ones (Madas et al. 2011; Madas and Balásházy 2011; Madas and Varga 2014). In fact, the numerical models are equivalent to configurations of spheres representing cell nuclei located in a rectangular cuboid corresponding to a small part of the epithelium in the large bronchi.

In case of basal (201 µm$^3$), ciliated (310 µm$^3$), goblet (243 µm$^3$), and other secretory cells (230 µm$^3$), volumes of cell nuclei is presented in Mercer et al. (1994). The volume of preciliated cell nuclei (310 µm$^3$) are supposed to be equal to the volume of ciliated cell nuclei. Cell nucleus volume of intermediate cells (156 µm$^3$) is computed by supposing that the nucleus/cytoplasm volume ratio of intermediate and basal cells are uniform. Cell volume of intermediate cells is quantified in the same way as in case of our earlier study (Madas and Balásházy 2011). The number of cell nuclei within a rectangular cuboid representing a small part of the epithelium is the product of the area of base of the cuboid (400 x 400 µm$^2$) and the cell numbers per unit basement membrane surface in the large bronchi derived from experimental data (Mercer et al. 1994).

Mercer et al. (1991) have determined the total cross section of nuclei of a given cell type relative to the cross section of the sample at seven different depths in the epithelium of the large bronchi. Continuous probability density functions for the depth of the centre of the cell nuclei were obtained by linear interpolation of the measured depth distributions. Location of the cell nuclei were determined by the following algorithm.

First, the depth of the centre of nucleus is selected randomly from the continuous depth distribution. If the distance between the centre of the nucleus and either the top or the bottom face of the cuboid is smaller than the radius of the nucleus, then the random selection will be repeated using the same depth distribution function. After the depth of the centre of the nucleus is determined, the first two coordinates are selected randomly from a uniform distribution over the area of base of the cuboid (400 x 400 µm$^2$). If the distance between the centre of the nucleus and one of the faces of the cuboid is smaller than the radius of the nucleus, then the selection of the first two coordinates, but not the depth will be repeated.

If the distance between the centre of the new nucleus and the centre of one of the previously located nuclei is smaller than the sum of the radii of the nuclei, the selection of the first two coordinates but not the depth will be repeated. The number of unsuccessful repetitions due to other nuclei is counted for a given nucleus, and if it reaches an arbitrary threshold[1], a new algorithm is applied. In this case, the centre of the new nucleus is shifted by applying the following equation:

$$C_{new}(i) = C_{old}(i) + (r_{old} + (1+(10-k)/10) \cdot r_{new}) \cdot u(i) \ , \quad (1)$$

where $C_{new}(i)$ and $C_{old}(i)$ are the *i*th coordinates of the centre, while $r_{new}$ and $r_{old}$ are the radii of the new and the previously selected nucleus intersected by the new nucleus, respectively. *k* is an integer variable running from 0 to 10, while *u(i)* is the *i*th coordinate of the vertical projection of a unit vector with a direction towards the centre of the new nucleus from the centre of the previously selected nucleus. If the shift is unsuccessful due to another nucleus, the value of *k* is increased by 1, and equation (1) is applied with the parameters characterizing the nucleus lastly intersected by the new nucleus.

If all the shifts are unsuccessful, i.e. there is an intersection with one of the previously selected nuclei for all values of *k* between 0 and 10, a new horizontal location is selected uniformly over the area of base of the epithelium model, and *k* is set to zero. Then location is searched around this new location by applying the shifting method. Finally, if an appropriate location is found without intersections with the previously selected nuclei and the faces of the epithelium model, the search for the position of the next nuclei is started until all the nuclei are located in the epithelium model.

---

1 This threshold is the ratio of the diameter of basal cell nucleus and the thickness of the epithelium multiplied by the total number of nuclei in the epithelium model (already located + to be located).

In order to study the microdosimetric consequences of hyperplasia, different epithelium models were generated by supposing different number of progenitor cells. For the sake of simplicity, only the number of basal cells have been increased, which are the main progenitor cells of the central airways (ICRP, 2015). The number of secretory cells (including neuroendocrine cells) was kept constant, although they are considered as progenitor cells (ICRP, 1994; 2015). Despite the evidence for the induction of goblet cell hyperplasia upon exposure to different drugs and irritants (Rogers 1994; 2003), the number of goblet cells and so the mucus thickness was not changed either.

We supposed that the additional basal cells have the same volume as the original ones (622.8 µm) (Mercer et al. 1994; Madas and Balásházy 2011). Thus the thickness of the tissue was increased accordingly from 57.8 µm characterizing the epithelium of the large bronchi in normal conditions (Mercer et al. 1991). The relative number of additional basal cells, the absolute number of basal cells per unit basement membrane surface, and the epithelium thickness for models with different measure of hyperplasia are summarized in Table 1.

**Table 1.** Numbers characterizing the measure of basal cell hyperplasia: increase in number of basal cells relative to their normal number, absolute number of basal cells per unit basement membrane surface, and the thickness of the epithelium.

| Relative increase in basal cell number (%) | Total basal cell number per unit surface (mm$^{-2}$) | Thickness of the epithelium (µm) |
|---|---|---|
| 0% | 17 100 | 57.80 |
| 5% | 17 956 | 58.33 |
| 10% | 18 813 | 58.86 |
| 50% | 25 650 | 63.12 |
| 100% | 34 200 | 68.45 |
| 150% | 42 750 | 73.77 |
| 200% | 51 300 | 79.10 |

The locations of the centres of cell nuclei are determined in the same way as for the epithelium with normal thickness. The only difference is that the probability density functions for basal cells are obtained by multiplying the depths where experimental data are available with the thickness ratio of the epithelium models. In this way, basal cells are placed deeper in the hyperplasia models than in the normal one, while the depth distributions of other cell nuclei change only due to the lack of reselection of those nuclei whose centres were placed too close to the bottom face of the cuboid representing the normal epithelium.

*Microdosimetry model*

It is supposed that an 11-micron-thick layer consisting of an upper mucus and a bottom cilia layer covers the epithelium (ICRP, 1994), and decays occur on the top of the mucus layer. Only alpha-decays were considered in this study. Since we focus on the deposition hot spots in the peak of the bifurcations, decays taking place on the surface of other parts of the bronchi can be neglected besides decays taking place on the modelled part of the epithelium. The decay ratio characterises the most exposed parts of the large bronchi of a worker in the former New Mexico uranium mine: 10.4% of the alpha-particles are emitted by $^{218}$Po (6.00 MeV) and 89.6% by $^{214}$Po (7.69 MeV) (Madas and Balásházy 2011).

Alpha particles are considered as straight lines, and their range was obtained with the freely available "SRIM" software (Ziegler et al. 2008). From the compound dictionary of the software, "trachea" was chosen as a target representing both the bronchial epithelium and the layers covering the epithelium. Cell nucleus hits and cell nucleus doses are determined by an own-developed Monte-Carlo code.

Based on our earlier study, 0.047 WLM (working level month)[2] equivalent to 8 h of work in a mine environment characterised by an exposure rate of 1 WL (working level) results in approximately 0.125 µm$^{-2}$ alpha-decay per unit surface area corresponding to 0.93 Gy tissue dose in a 0.14 mm$^2$ hot spot (Szőke et al. 2008; Madas and Balásházy 2011) if mucociliary clearance is not considered. Mucociliary clearance would decrease the number of alpha-decays and tissue dose in the hot spot by approximately one order of magnitude (Farkas and Szőke 2013; Madas 2016).

The tissue dose of 0.93 Gy (for 1 WLM exposure) applies only to normal conditions, when the tissue thickness is 57.8 µm. For hyperplastic conditions, the tissue thickness is higher, and the tissue dose can be different. For this reason, the independent variable in our simulations is the number of alpha-decays per unit surface. Since it is proportional to exposure in WLM independently on the tissue thickness, the exposure in WLM is also depicted in figures 1-3.

In order to estimate empirical standard deviations, 500 numerical epithelium models were generated for each given measure of hyperplasia, and 500 independent simulations were performed using these 500 different configurations of cell nuclei for each number of alpha-decays per unit surface area.

## Results

Figure 1 shows that the induction of hyperplasia results in changes in the average tissue dose. Due to the additional number of basal cells, tissue thickness increases, and the dose absorbed by the epithelium with the same surface area decreases. The left panel shows the average tissue dose as the function of alpha decays per unit surface area for different number of additional basal cells. Tissue dose increases linearly by the number of alpha-decays. The slope of the curve decreases in the sequence of 5.98 Gyµm$^2$ (0%), 5.92 Gyµm$^2$ (5%, not shown), 5.87 Gyµm$^2$ (10%, not shown), 5.47 Gyµm$^2$ (50%) 5.05 Gyµm$^2$ (100%), 4.68 Gyµm$^2$ (150%), and 4.37 Gyµm$^2$ (200%). It means that doubling the number of basal cells from their normal number reduces tissue dose by 15.6%, while tripling it results in a decrease in tissue dose of about 26.9%.

The reduction in tissue dose can also be seen in the right panel, where tissue dose is plotted for 8 h of work in a mine environment as the function of the measure of hyperplasia, i.e. the relative increase in basal cells number. The different curves refer to different exposure rates.

As the simulations were performed for the deposition hot spots, even relatively small exposure rates result in high local doses. One working level month results in 12 mSv effective dose if the detriment-adjusted nominal risk coefficient for radon exposure of 5 x 10$^{-4}$ WLM$^{-1}$ (ICRP, 2010) and the value of total radiation detriment per unit effective dose of 4.2 x 10$^{-5}$ mSv$^{-1}$ for workers (ICRP, 2007) are considered. An exposure rate of 0.518 WL results in 6.22 WLM/year equivalent to 74 mSv/year effective dose rate, but causes a local tissue dose rate of almost 0.4 Gy/day which is reduced to less than 0.35 Gy/day if basal cell number is tripled.

---

2  Working level month is the historical unit of exposure to radon progeny applied to the uranium mining environment. One WLM (equivalent to 3.54 x 10$^{-3}$ Jh m$^{-3}$) is defined as the cumulative exposure from breathing an atmosphere at a concentration of one working level (WL) for a working month of 170 h. A concentration of 1 WL is any combination of short-lived radon progeny in one litre of air that will result in the emission of 1.3 x 10$^5$ MeV of alpha-energy (ICRP, 2010).

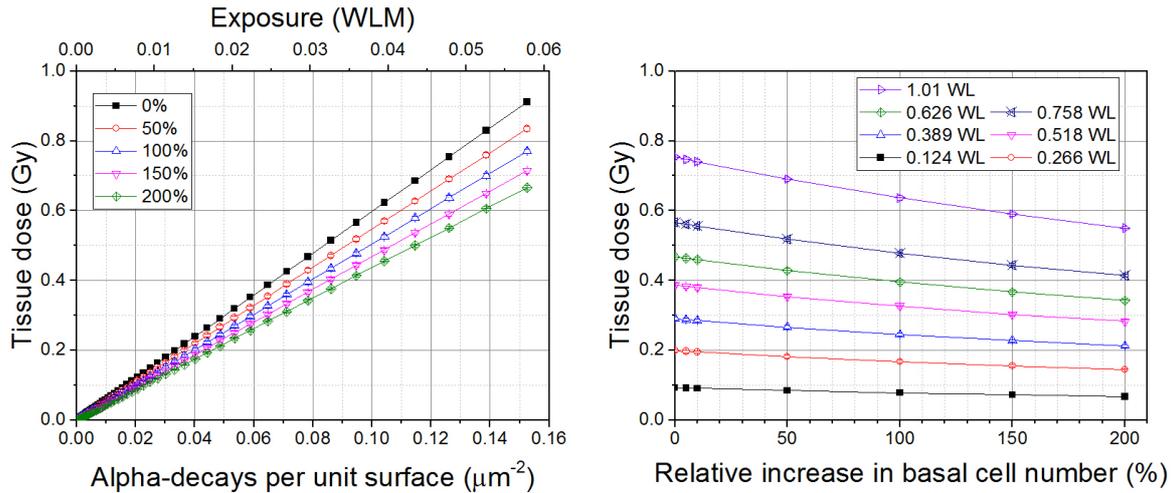

*Figure 1: Tissue dose as the function of the number of alpha-decays per unit surface (left panel) and tissue dose as the function of the measure of basal cell hyperplasia (right panel) if 8 h is spent in a mine environment. In the left panel, different curves refer to different number of additional basal cells relative to their normal number. In the right panel, different curves refer to different exposure rates. All date applies to the aerosol deposition hot spots in the large bronchi, where even small exposure rates result in high local doses. In case of exposure in WLM and exposure rate in WL, mucociliary clearance is neglected.*

As ICRP (1994) identifies radiosensitive target cells in the human respiratory tract, the radiation burden of these cells is of higher interest than tissue dose. ICRP (1994) considers basal cells and columnar secretory cells as progenitors of the bronchial epithelium. As *in vitro* experiments show that cell survival probability depends on cell nucleus hits (Hei et al. 1997; Soyland and Hassfjell 2000), Figure 2 shows the average number of hits of basal (left panel) and secretory cells (right panel) identified with "other secretory cells" of the large bronchi in Mercer et al. (1991; 1994).

The average number of hits increases linearly by the number of alpha-decays per unit surface. The additional number of basal cells decrease the slope of the curves for basal cells in the sequence of 5.08 $\mu m^2$ (0%), 4.94 $\mu m^2$ (5%, not shown), 4.81 $\mu m^2$ (10%), 3.73 $\mu m^2$ (50%), 2.57 $\mu m^2$ (100%), 1.72 $\mu m^2$ (150%), and 1.16 $\mu m^2$ (200%). The relative difference in the number of hits is higher than it is in tissue dose: doubling the number of basal cells from their normal number reduces the average number of hits by 49.4%, while tripling the number of basal cells results in a decrease of average hit number of about 77.2%.

This decrease in the average number of basal cell nucleus hits is the consequence of the different depth distribution of basal cells due to the increased epithelial thickness. In the same time, the average number of secretory cell nucleus hits does not change by the measure of basal cell hyperplasia, because their depth distribution is not affected by the additional basal cells and increased tissue thickness. The slope of the curves in the right panel varies minimally between 15.6 and 15.8 $\mu m^2$.

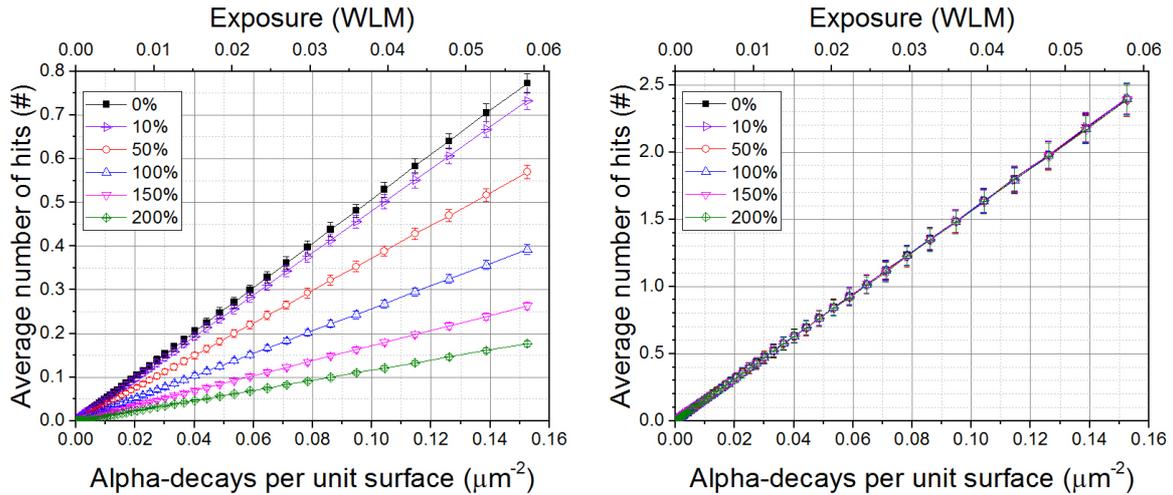

*Figure 2: Average number of cell nucleus hits for basal (left panel) and secretory cells (right panel). The different curves refer to the additional number of basal cells relative to their normal number. In case of exposure in WLM, mucociliary clearance is neglected.*

The average dose of progenitor cell nuclei was also computed. Results for basal and secretory cell nuclei are plotted in Figure 3 (left and right panel, respectively). Similarly to the average tissue dose and the average number of nucleus hits, the average cell nucleus dose also increases linearly by the number of alpha decays per unit surface. For secretory cells, the slope of curve is independent on the number of basal cells: it varies between 6.22 and 6.28 Gyµm$^2$.

For basal cells, the slope decreases by tissue thickness in the sequence of 2.59 Gyµm$^2$ (0%), 2.51 Gyµm$^2$ (5%, not shown), 2.44 Gyµm$^2$ (10%), 1.85 Gyµm$^2$ (50%), 1.25 Gyµm$^2$ (100%), 0.829 Gyµm$^2$ (150%), and 0.554 Gyµm$^2$ (200%). The relative reduction in average dose of basal cell nuclei is similar to the average hit number, but higher than the average tissue dose. Doubling the number of basal cells from their normal number reduces the average dose of nuclei by 51.7%, while tripling the number of basal cells results in a decrease of about 78.6% of average basal cell nucleus dose.

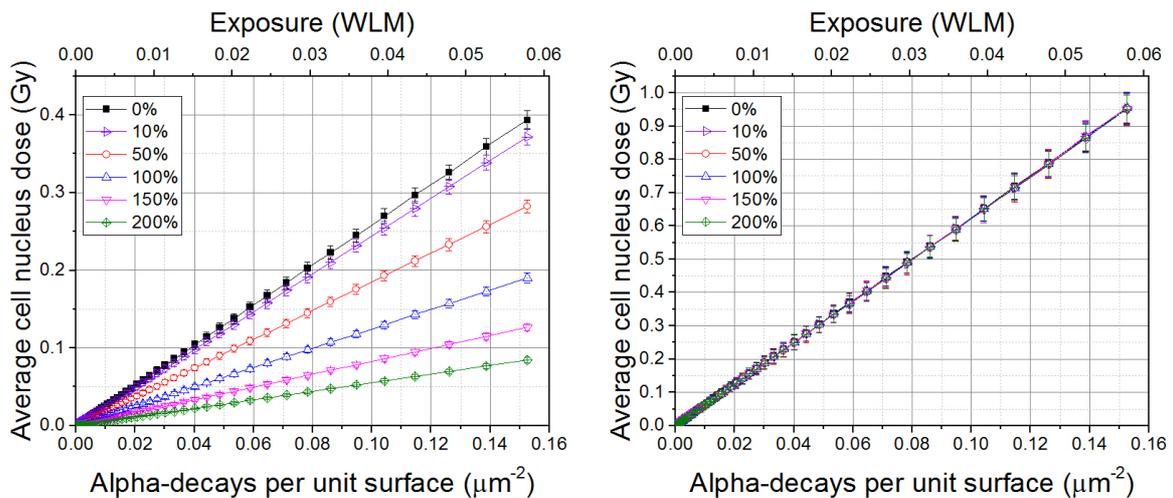

*Figure 3: Average cell nucleus dose for basal (left panel) and secretory cells (right panel). The different curves refer to the additional number of basal cells relative to their normal number. In case of exposure in WLM, mucociliary clearance is neglected.*

Figure 4 shows the hit distribution of basal and secretory cells in a deposition hot spot if a uranium miner works eight hours in an exposure rate of 1.01 WL resulting in approximately 0.125 µm$^{-2}$ alpha-decay per unit surface area if mucociliary clearance is neglected. It reveals that the changes in average hit number and average dose of basal cell nuclei do not show the essential microdosimetric consequences of the induction of hyperplasia. While the number of basal cells receiving 1, 2, 3, and 4 hits are significantly lower for the epithelium with tripled basal cell number (green bars with horizontal pattern) than for the epithelium having doubled or less number of basal cells (left panel), it cannot explain the decrease in the average hit number of basal cells. Instead, the number of non-hit basal cells increases dramatically due to the induction of hyperplasia, which results in the strong reduction of average hit number of basal cell nuclei. The hit distribution of secretory cell nuclei (right panel) is not affected by the higher basal cell number and the increased epithelium thickness.

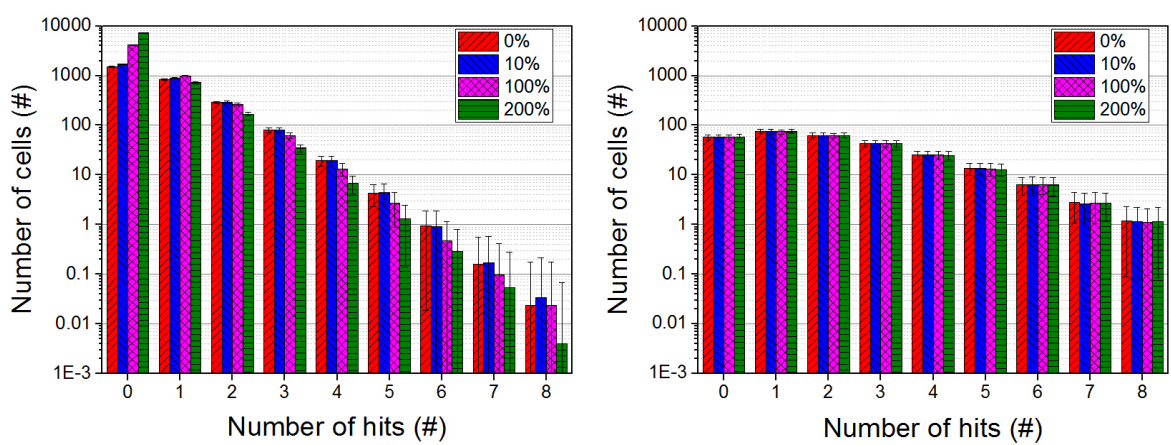

*Figure 4: Hit distribution of basal (left panel) and secretory cell nuclei (right panel) in the deposition hot spot if a uranium miner works eight hours in an exposure rate of 1.01 WL for different measures of basal cell hyperplasia. Bars with different colours and patterns refer to the additional number of basal cells relative to their normal number. Mucociliary clearance is neglected.*

Figure 5 shows the dose distribution of nucleus of basal and secretory cells in a deposition hot spot if a uranium miner works eight hours in an exposure rate of 1.01 WL resulting in approximately 0.125 µm$^{-2}$ alpha-decay per unit surface area if mucociliary clearance is neglected. While the dose distribution of secretory cells (right panel) does not change due to the additional number of basal cells, the dose distribution of basal cells (left panel) shows the same increase in the non-hit fraction as Figure 4. It can also be seen that tripling the basal cell number results in an absolute reduction of cell receiving higher doses than 0.5 Gy.

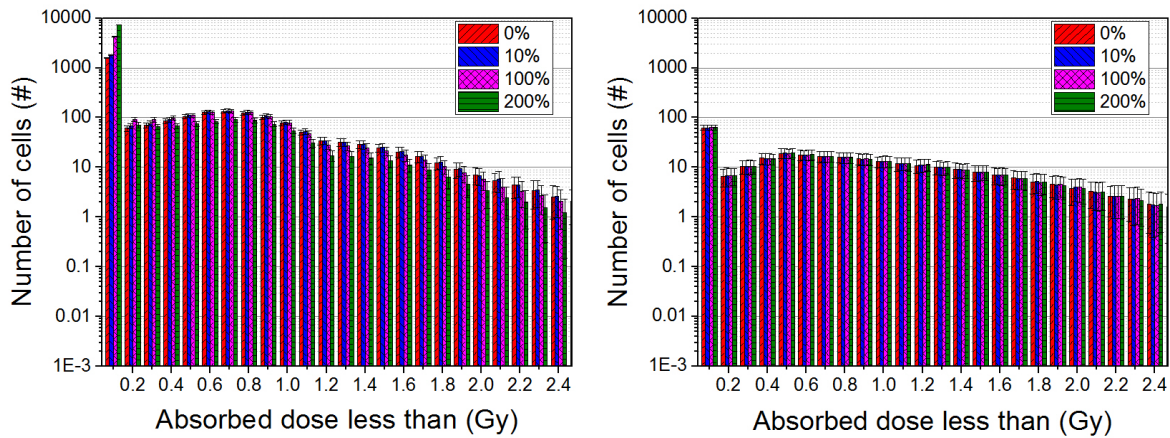

*Figure 5: Dose distribution of basal (left panel) and secretory cell nuclei (right panel) in the deposition hot spot if a uranium miner works eight hours in an exposure rate of 1.01 WL for different measures of basal cell hyperplasia. Bars with different colours and patterns refer to the additional number of basal cells relative to their normal number. Mucociliary clearance is neglected.*

## Discussion

*Previous exposures modulate the microscopic effects of a present exposure*

Earlier, the induction of progenitor cell hyperplasia in the deposition hot spots was put forward as a consequence of high cell death rate upon chronic exposure to radon progeny (Madas and Balásházy 2011). This study highlights that the induction of hyperplasia during previous exposures modulates the effects of a present exposure. In this work, we focussed on the modulation of microscopic doses, i.e. on the changes in hits received and doses absorbed by progenitor cell nuclei. We found that the increase in basal cell number and the resultant increase in tissue thickness decreases significantly the average dose and hit number of basal cell nuclei by strongly enhancing the non-hit fraction of basal cells. While the modulation of biological effects due to the induction of hyperplasia was not studied here, the fraction of basal cells shielded from radiation may have significant impacts on such biological effects of radiation as the changes in cell division rate of progenitors (Madas and Balásházy 2011) or the rate of clonal growth of preneoplastic cells (Heidenreich and Paretzke 2008).

Aiming the support of the system of radiological protection, significant efforts have made to develop biokinetic and dosimetric models taking into account the location of incorporated radionuclides and radiosensitive target cells in order to determine effective doses (ICRP, 1994; ICRP, 2006). However, even if these models were fully appropriate for acute exposures, the present study shows that the location of radiosensitive target cells may change due to radiation exposure. Thus dosimetry models considering the irradiation geometry characteristic of normal conditions may be inappropriate for dose estimation in case of chronic exposures. As internal exposures are frequently chronic exposures, it may be highly relevant for other incorporated radionuclides emitting short range particles.

*Hyperplasia induced by radon exposure as a radioadaptive response at the tissue level*

Radioadaptive responses belonging to the group of non-targeted effects are described as the reduced damaging effect of a challenging radiation dose when induced by a previous low priming dose (Tapio and Jacob 2007). The induction of hyperplasia has a similar effect reducing the local radiation burden of progenitor cells because of the increased tissue thickness and the increased depth of basal cell nuclei. From this point of view, basal cell hyperplasia induced by radon exposure can be reckoned as a radioadaptive response at the tissue level. Indeed, the induction of hyperplasia is considered as an adaptation, however, it is not mentioned among the radioadaptive responses.

Besides the similarities, however, there are some important differences between classic adaptive responses and hyperplasia induced by radon exposure. While most forms of adaptive responses manifest themselves at the cellular level, hyperplasia is an adaptation of the tissue. Another difference is that in case of hyperplasia, the reduction in biological damage is the result of the decreased local dose consequences of a given macroscopic exposure, while in case of classic adaptive responses, the damage caused by the same cellular dose is reduced. As opposed to classic adaptive responses, the priming dose for the induction of hyperplasia is not necessarily low, moreover high dose rates may be even more effective. Nevertheless, intercellular communication is a prerequisite for the induction of hyperplasia which is also part of the process of classic radioadaptive responses (Tapio and Jacob 2007).

While the induction of hyperplasia diminishes the local tissue damage by alpha-particles, on one hand, it may increase the risk of stochastic effects, on the other hand. Some analyses of epidemiological data suggests that radon exposure is primarily a promoting agent for lung carcinogenesis (Luebeck et al. 1999; Eidemüller et al. 2012). It is expected that the induction of progenitor cell hyperplasia is accompanied by an increase in the number of preneoplastic progenitor cells, and so it may contribute to promotion. It means that the induction of hyperplasia as an adaptation of the tissue does not necessarily lead to a decrease in cancer risk.

*Acute and chronic exposures, the dose dose rate effectiveness factor*

The role of dose rate as a modifier of radiation effects is reflected in the present system of radiological protection by the dose dose rate effectiveness factor (ICRP, 2007). It implies the assumption that the risk of the same dose is higher if the exposure is protracted. Challenging this assumption, however, a comparison of data from epidemiological studies involving low and high dose rates of low-LET radiation does not suggest any dose-rate effect (Jacob et al. 2009). In addition, epidemiology of lung cancer among uranium miners suggests directly the opposite, i.e. an inverse exposure rate effect (Lubin et al. 1995; Tomasek et al. 2008; Walsh et al. 2010). Among other, these studies feed the on-going debate about the role of dose rate as a modifier of radiation effects (Rühm et al. 2015)

The induction of hyperplasia provides an example for the basic differences between the effects of acute and chronic exposures. As progenitor cell hyperplasia contributes to the local tissue maintenance, it is also expected that its measure, i.e. the additional number of progenitors is dependent on cell death rate and so on dose rate. Therefore the induction of hyperplasia challenges the application of dose dose rate effectiveness factor from a mechanistic point of view because the effects of chronic exposures seem too complex to be described by a single value.

Quantitative studies on the effects of hyperplasia may help us to understand the biological effects of chronic exposures. Based on the equilibrium between cell death and cell proliferation, the measure of hyperplasia and division rate of progenitors can be estimated by quantifying cell death rate. These estimations can provide input for mathematical models analysing epidemiological data resulting in more accurate risk assessment for chronic exposures to radon progeny. While it is not clear whether the induction of hyperplasia increases or decreases cancer risk, it may highly modulate both the biological and health effects of radon exposure and other incorporated radionuclides.

## Conclusions

In the present study, it was pointed out that recent exposure history of the tissue modulating its geometrical properties determines the microscopic dose consequences of a given macroscopic exposure to radon progeny. As the local tissue dose and the radiation burden of basal cells decreases due to the induction of hyperplasia, it can be considered as a radioadaptive response which manifests itself at the tissue level. Such an adaptation of the tissue provides an example of the fundamental differences between acute and chronic exposures, and challenges the validity of the application of the dose dose rate effectiveness factor from a mechanistic point of view. In addition, as the location of radiosensitive target cells may change due to radiation exposure, dosimetry models considering the irradiation geometry characteristic of normal conditions may be inappropriate for dose estimation in case of chronic exposures. As internal exposures are frequently chronic, such changes in tissue geometry may be highly relevant for other incorporated radionuclides emitting short range particles.

## Acknowledgements

The author thank Kornél Fél for his help in running simulations in Ubuntu Linux environment. This work was supported by the European Union and the State of Hungary, co-financed by the European Social Fund in the framework of TÁMOP 4.2.4. A/2-11-1-2012-0001 'National Excellence Program' (A2-EPFK-13-0160) and by the National Research, Development and Innovation Office under the contract VKSZ_14-1-2015-0021.